# Influence of multiorbitals and anisotropic Coulomb interactions on isotope effect coefficient in doped Fe-based superconductors

Luxmi Rani[1,2*] and Ajay[3]

[1]*Theoretical Physics Division, Physical Research Laboratory, Ahmedabad - 380009, India*
[2]*Department of Applied Sciences and Engineering, Indian Institute of Technology Roorkee, Saharanpur Campus, Saharanpur- 247001, India.*
[3]*Department of Physics, Indian Institute of Technology Roorkee - 247667, India*
***Email**: Luxmiphyiitr@gmail.com, luxmi@prl.res.in*

## Abstract

The present work describes the theoretical analysis of isotope effect coefficient as a function of transition temperature in two orbital per site model Hamiltonian in iron based superconducting system. The expression of isotope effect coefficient has been computed numerically and self-consistently by employing Green's function technique within the BCS- mean-field approximation. It is observed that the isotope effect coefficient increases with the increase of the hybridization while with the increase in Coulomb interaction it starts decreasing. On increasing the carrier density per site in two orbital per site iron pnictide system, isotope effect coefficient ($\alpha$) exhibits large values (much higher than BCS limit) at lower temperatures. While in the underdoped case, isotope effect coefficient shows minimum value in superconducting states of the iron based systems. Furthermore, it has been found that the large value of the isotope effect coefficient is the indication of the fact that the contribution of phonon alone is inadequate as the origin of superconductivity in these systems. Finally, the obtained theoretical results have been compared with experimental and existing theoretical observations in iron based superconductors.

**Keywords:** Iron based superconductors, Isotope effect coefficient, and Green's function technique.

**PACS numbers:** 74.70.Xa, 74.20.Fg

# Introduction

The discovery of isotope effect by Fröhlich (1950) [1] provided a basis for the microscopic theory called BCS-theory of superconductivity. The essence of BCS theory lies in its microscopic mechanism that provides attractive interaction between two electrons of opposite spin and momentum through lattice vibrations (phonon). Due to attractive interaction, these pairs electrons form a bound state called Cooper pairs and give rise to the superconducting state. Taking into account the theory of superconductivity based on lattice vibration, Maxwell [2] and Reynolds [3] studied experimentally the variation of superconducting transition temperature with the isotope substitution. Thereafter, the BCS theory [4] supported the concept of electron–phonon interaction, proposed by Bardeen, Cooper, and Schrieffer in 1957, and explains the origin of superconductivity in conventional superconductors, though the origin of superconductivity in recent Iron and Cuprate high $T_c$ systems still lacks proper understanding. The superconducting transition temperature ($T_c$) of phonon- mediated superconductors is usually low (eg. Hg, Pb etc.) with isotope exponent α= 0.5. The BCS theory predicts that the superconducting transition temperature ($T_c$) to be proportional to the Debye frequency ($\omega_D$) and hence to the square root of isotope mass M (*i.e.* $T_c \propto M^{-\alpha} \propto \omega_D$, where α is the isotope effect coefficient). Therefore the isotope effect coefficient is defined as;

$$\alpha = -\partial \ln T_c / \partial \ln M .$$

For conventional superconductors, the exact BCS value of the isotope effect coefficient is 0.5.

The discovery of high -$T_c$ superconductivity in cuprates by Bednorz and Müller [5] and the recent finding of the high -$T_c$ iron based superconductors by Kamihara group in 2008 [6] changed the traditional concept and clearly indicated that BCS theory based on the electron-phonon interaction may not be able to explain such high $T_c$'s and spin fluctuation since antiferromagnetic background can also contribute to the pairing mechanism, and still a debatable issue from the theoretical point of view. The isotope effect coefficients show a deviation (above and below the BCS limit) in iron based high-$T_c$ superconductors and need careful attention in any theoretical analysis. Qui *et al.* [7] studied neutron scattering measurements in LaO$_{0.87}$F$_{0.13}$FeAs that indicates the existence of spin density wave (SDW) orders in the normal and superconducting state. They have also observed that the BCS theory based on electron –phonon interaction fails to produce high $T_c$ (26 K) for most of iron based superconducting systems. Further, Boeri *et al.* [8, 9] reported the weak electron- phonon coupling in iron oxypnictide LaFeAs(O, F) system using density functional perturbation theory and predicted that the electron-phonon interaction alone will be insufficient to achieve high-$T_c$ superconductivity in iron based superconductors. However, Bang *et al.* [10] studied the isotope effect in Fe based system and found a large value of the iron isotope coefficient $\alpha_{Fe} = 0.4$.

In the year 2009, Liu *et al.* [11] studied the isotope effect in two doped iron based systems SmFeAsO$_{1-x}$F$_x$ ($T_c$= 38K) and Ba$_{1-x}$K$_x$Fe$_2$As$_2$ by inducing the oxygen $^{18}$O in place of $^{16}$O (oxygen isotope effect) and substituting (iron isotope) $^{54}$Fe in place of $^{56}$Fe. They found that iron isotope effect much larger than the oxygen isotope in the iron pnictide system. The author suggested that conducting layer of the Fe-As plane may be responsible for evolving superconductivity in these systems. On the other hand, the first evidence of inverse isotope effect in high $T_c$ superconductors (Ba, K) Fe$_2$As$_2$ has been found by Shirage *et al.* [12] and affirmed that the superconducting mechanism in iron based superconductors may not be clearly understood by the conventional BCS theory. Yanagisawa *et al.* [13] studied the inverse isotope effect report [12] by taking into account multiband BCS superconductivity in (Ba, K) Fe$_2$As$_2$ systems. Thereafter, Holder and Keller [14] have commented on multiband model [13] and observed that there is no sign change in the isotope exponent. Therefore, the isotope effect in multiorbital iron pnictide systems has not been clearly understood from the theoretical point of view and need attention.

On the basis of First principal approach, Boeri *et al.* [8, 9] found that the undoped compound (LaOFeAs) shows a sharp spectral peak in the density of states spectrum due to magnetic instability while the introduction of electrons suppresses the electronic states at the Fermi level, without showing significance in the electronic band structure. These authors further pointed out a very small electron –phonon coupling change. A recent experimental study [15] also has been suggested that the electron –phonon coupling play an important role in pairing mechanism of iron based superconductors causing iron isotope effect. Also, the dependence of transition temperature ($T_c$) on the existence of FeAs$_4$ tetrahedral [16, 17] has been observed. Generally, Iron based superconductors have a metallic parent compound and show small electronic anisotropy [18]. The electronic band structure and Fermi surface of these materials contain two hole pockets at the centre (i.e. Γ (0, 0)) point of Brillouin zone (BZ) and two electron pockets at the BZ corners (i.e. X (±π, 0) or (0, ±π)) points [19, 20, 21] and exhibit extended s-wave pairing symmetry [22]. The Spin fluctuation mechanism for mediating pairing in iron based superconductor (Fe-BSC) has also been proposed [10] and a strong magnetic moment at the Fe atomic site is obtained. Further, Yin *et al.* [23] predicted the $s_{\pm}$-wave pairing in orbital space which arises due to the electron correlations. Recently [24], it has been emphasized that spin and lattice degrees of freedom may cooperate and simultaneously give rise pairing in iron based superconductors.

A clear understanding of isotope effect in iron based superconductors may pinpoint the origin of superconductivity in iron based materials. The several recent experimental data related to isotope effect in iron based superconductors [10, 11, 25] have been reported as positive Fe isotope effect (Fe-IE) exponent $\alpha_{Fe}$=0.34 in $Ba_{0.6}K_{0.4}Fe_2As_2$ and in $SmFeAsO_{0.85}F_{0.15}$ $\alpha_{Fe}$ is 0.37 [11], while in $FeSe_{1-x}$ [25] the value of $\alpha_{Fe}$ = 0.81, respectively. In contrast, Shirage *et al.* [26,27] have reported a negative $\alpha_{Fe}$= −0.18 and −0.024 isotope coefficient for $Ba_{0.6}K_{0.4}Fe_2As_2$ and $SmFeAsO_{1-y}$ materials, respectively. Very recently, Tsuge *et al.* [28] reported the iron isotope exponent $\alpha_{Fe}$=−0.19 in $(Ca_{0.4}Na_{0.6})_2Fe_2As_2$ system [28]. Thereafter, Khasanov *et al.* [29], and Kim and Granath [30] provided a brief report on structural changes and variation in lattice parameters due to substitution of iron isotopes in different types of iron based superconductors. Also, they predicted that a large isotope shift in $T_c$ can be found through the little change in lattice via isotope substitution in these systems. Further, Choi et al. [31] reported both positive and inverse isotope effect in these systems. Recently, Abah *et al.* [32] theoretically analyzed the π-phase shifted s-wave in iron based superconductors and also found that the thought of both positive and negative isotope effect is applicable for these systems. Chu *et al.* [33] first time presented an experimental link between iron isotope exponent and $T_c$ and observed the microscopic mechanism of superconductivity in these systems. As of now, the theoretical issues related to isotope effect on $T_c$ and hence the microscopic origin of superconductivity has not been clearly understood so far in these systems.

Till date, several theoretical studies have been proposed two orbital [20, 34], three orbital [35] and five orbital [19] models based calculations. Further, the importance of inter and intra orbital Coulomb interactions and Hund's coupling has been emphasized, due to the tendency of electrons occupation in different orbitals and in the high spin state, in these systems [36]. Recently, we have theoretically analyzed the electronic spectra within two orbital per site and three orbital per site tight binding model for iron based superconductors. It is pointed out that the Hund's coupling accumulate the spectral weight (i.e. electronic states) near to Fermi level while Coulomb interactions suppress the electronic states close to Fermi level and may play a role in stabilizing superconducting state [37-39] in these systems. As an addition of these studies, the interplay of on-site Coulomb interactions, multiorbital coupling, and doping on the isotope effect need to be understood properly.

Therefore, in the light of above facts, it is important to analyze the impact of Fe $3d$ multi-orbital's coupling, on-site electronic Coulomb interactions, Hund's coupling term, and doping on the isotope effect coefficient ($\alpha$). For this purpose, we have performed numerical computation within the two orbital per site model Hamiltonian and employ Green's function formalism within BCS-mean field approximation to maintain the self-consistency. The numerically computed isotope effect data are compared with theoretical results and existing experimental investigations related to these systems. In the next section, the theoretical formalism of superconducting order parameter and isotope effect coefficient as a function of the superconducting transition temperature ($T_c$) within the two orbital per site model for iron pnictides is presented.

**Model and Formulation of Isotope Effect Coefficient ($\alpha$):**

We have focused on 1111 family of iron based superconductor, where LaO and FeAs layers are stacked alternately in two dimensional (2D) square lattice. The Fe atoms are in a fourfold coordination having shape of a $FeAs_4$-tetrahedron. The Fe atoms form square net with a Fe-Fe distance of approximately $2.8 A°$ [40] and the As-atoms are located above and below the Fe-plane as shown schematically in figure 1. We have considered two orbital model characterized by $d_{xz}$ and $d_{yz}$ orbitals per Fe site. Both orbitals $d_{xz}$ and $d_{yz}$ have a dominant contribution to the electronic states near the Fermi surface ($E_F$) in comparison to other orbitals [20] and strongly influence the superconducting properties. The hopping between $d_{xz}$ and $d_{yz}$ orbitals in the two band model is illustrated in (figure 1). The $t_1$ is the nearest neighbor hopping between $\sigma$-orbitals, and $t_2$ is the nearest neighbor hopping between $\pi$-orbitals. The $t_3$ is the next nearest neighbor hopping between similar orbitals and $t_4$ is the next nearest neighbor hopping between different orbitals. [See figure 1]

Initially, one need to develop the kinetic part of tight binding model Hamiltonian of iron pnictide based on two orbitals ($d_{xz}$ and $d_{yz}$) per site on two dimensional square lattice of iron [37] in momentum space. This may be written as follows:

$$H_0 = \sum_{k,\sigma} \Psi^+_{k,\sigma} \left\{ (\varepsilon_+(k) - \mu) I + \varepsilon_-(k)\tau_3 + \varepsilon_{xy}(k)\tau_1 \right\} \Psi_{k,\sigma}. \tag{1}$$

Where, μ is the chemical potential, and the wave function $\Psi_{k,\sigma}$ is given by

$$\Psi_{k,\sigma} = \begin{pmatrix} C_{d_{yz},k,\sigma} \\ C_{d_{xz},k,\sigma} \end{pmatrix}.$$

$\Psi_{k,\sigma}$ is the annihilation operator for spin σ electrons in the two orbital and similarly $\Psi^+_{k,\sigma}$ is the creation operator. **I** is 2×2 identity matrix, $\tau_1$ and $\tau_3$ are the standard Pauli's matrices given by $\tau_1 = \begin{pmatrix} 0 & 1 \\ 1 & 0 \end{pmatrix}$, $\tau_3 = \begin{pmatrix} 1 & 0 \\ 0 & -1 \end{pmatrix}$, and $I = \begin{pmatrix} 1 & 0 \\ 0 & 1 \end{pmatrix}$ respectively, also

$$\varepsilon_\pm(k) = \frac{\varepsilon_x(k) \pm \varepsilon_y(k)}{2}.$$

$$\varepsilon_x(k) = -2t_1 \cos k_x - 2t_2 \cos k_y - 4t_3 \cos k_x \cos k_y, \tag{1a}$$

$$\varepsilon_y(k) = -2t_2 \cos k_x - 2t_1 \cos k_y - 4t_3 \cos k_x \cos k_y, \quad \text{and} \tag{1b}$$

$$\varepsilon_{xy}(k) = -4t_4 \sin k_x \sin k_y. \tag{1c}$$

Here, $t_1$, $t_2$, $t_3$ and $t_4$ are nearest neighbor and next nearest neighbor hopping energies between two orbital's $d_{xz}$ and $d_{yz}$ per site (as shown in Figure 1) on a square FeAs-lattice in iron based superconductors. Thus, the kinetic part of Hamiltonian can be rewritten as follows:

$$H_0 = \sum_{k,\sigma} \begin{pmatrix} C^+_{d_{yz},k,\sigma} & C^+_{d_{xz},k,\sigma} \end{pmatrix} \begin{pmatrix} \varepsilon_+(k) + \varepsilon_-(k) - \mu & \varepsilon_{xy}(k) \\ \varepsilon_{xy}(k) & \varepsilon_+(k) - \varepsilon_-(k) - \mu \end{pmatrix} \begin{pmatrix} C_{d_{yz},k,\sigma} \\ C_{d_{xz},k,\sigma} \end{pmatrix} \tag{2}$$

To solve above equation, one can employ Bogoliubov transformation [37] and therefore, the kinetic part of model Hamiltonian possess two bands in FeAs- square lattice given as follows:

$$H_0 = \sum_{k,\sigma} \left\{ \varepsilon_+(k) - \mu + \sqrt{(\varepsilon_-^2(k) + \varepsilon_{xy}^2(k))} \right\} C^+_{1k,\sigma} C_{1k,\sigma} + \sum_{k,\sigma} \left\{ \varepsilon_+(k) - \mu - \sqrt{(\varepsilon_-^2(k) + \varepsilon_{xy}^2(k))} \right\} C^+_{2k,\sigma} C_{2k,\sigma}. \tag{3}$$

For simplicity, it may be written as:

$$H_0 = \sum_{k,\sigma} \varepsilon_{1k} C^+_{1k,\sigma} C_{1k,\sigma} + \sum_{k,\sigma} \varepsilon_{2k} C^+_{2k,\sigma} C_{2k,\sigma} = \sum_{r=1,k,\sigma}^{2} \varepsilon_{rk} C^+_{rk,\sigma} C_{rk,\sigma} \quad . \tag{4}$$

Where quasi-particles energy dispersion of each band is described as follows [20,37]:

$$\varepsilon_{12k} = \left\{ \varepsilon_+(k) - \mu \pm \sqrt{(\varepsilon_-^2(k) + \varepsilon_{xy}^2(k))} \right\} \tag{5}$$

Now, we assume that superconducting correlations exist in both the bands, and as a interaction component; BCS effective attractive interaction and coupling between the bands along with intra and inter band interactions, and Hund's coupling term need to be included in model Hamiltonian. Thus, one can write the model Hamiltonian for pnictide superconductors within two band scenario [37] in momentum space as follows:

$$H = H_0 + H_{Intra} + H_{Inter}. \tag{6}$$

Where,

$$H_0 = \sum_{r,k,\sigma} (\varepsilon_{rk} - \mu) C^+_{rk\sigma} C_{rk\sigma},$$

$$H_{Intra} = \sum_{r,kk',\sigma} U_{kk'} C^+_{rk\sigma} C^+_{r-k-\sigma} C_{rk'\sigma} C_{r-k'-\sigma} + U' \sum_{r,kk',\sigma} n_{rk\sigma} n_{rk'-\sigma},$$

$$\begin{aligned} H_{Inter} &= \sum_{\substack{r \neq s, \\ k,\sigma}} \varepsilon_{rsk} (C^+_{rk\sigma} C_{sk\sigma} + C^+_{sk\sigma} C_{rk\sigma}) + U'' \sum_{\substack{r \neq s, \\ kk',\sigma}} n_{rk\sigma} n_{sk'-\sigma} \\ &+ J_H \sum_{r \neq s} \sum_{kk',\sigma} (C^+_{rk\sigma} C^+_{r-k-\sigma} C_{s-k'-\sigma} C_{sk'\sigma} + hc) \end{aligned}.$$

Where, r and s denotes two bands (1 and 2), and h.c means Hermitian conjugate. The $n_{rk\sigma} = C^+_{rk\sigma} C_{rk\sigma}$, and $n_{rk-\sigma} = C^+_{rk-\sigma} C_{rk-\sigma}$, represents the occupation number of the band electronic states of spin σ and -σ electrons. $C_{rk\sigma} (C^+_{rk\sigma})$ is the annihilation (creation) operator of electrons with spin σ (↑, ↓) and momentum (k) in each band and μ is the chemical potential. Here, $U_{kk'}$ is the effective BCS attractive interaction within each band and we have not gone into the detail microscopic origin of nature of attractive interaction forming Cooper pairs and superconductivity in these systems. The $\varepsilon_{rsk}$ is the term taking care of hybridization of electronic states of two bands. U' and U" are intra-band and inter-band Coulomb repulsions, which described the on-site electronic interaction, respectively. In the present work, we have taken contribution of onsite intra and inter orbital on same footings. Therefore, we have only treated onsite Coulomb interaction part between electrons of opposite spin in different bands. The term $J_H$ represents Hund's coupling energy in Fe $3d$ orbitals. In order to understand the microscopic mechanism of superconductivity in Fe-based materials, the isotope effect is an important parameter. Therefore, we have attempted the isotope effect analysis as a function of superconducting transition temperature within two orbital per site model Hamiltonian for iron pnictides. Also, we employ the equation of motion approach and derived four coupled Green's function equations within BCS mean field approximation as follows:

$$\{\omega - (\varepsilon_{1k} - \mu + U'\langle n_1 \rangle + U''\langle n_2 \rangle)\}\langle\langle C_{1k\sigma} | C^+_{1k\sigma} \rangle\rangle_\omega = \frac{1}{2\pi} + \varepsilon_{12k} \langle\langle C_{2k\sigma} | C^+_{1k\sigma} \rangle\rangle_\omega + (\Delta_{k11} + \Delta_{H22})\langle\langle C^+_{1-k-\sigma} | C^+_{1k\sigma} \rangle\rangle_\omega, \qquad (7)$$

$$\{\omega + (\varepsilon_{1k} - \mu + U'\langle n_1 \rangle + U''\langle n_2 \rangle)\}\langle\langle C^+_{1-k-\sigma} | C^+_{1k\sigma} \rangle\rangle_\omega = -\varepsilon_{12k} \langle\langle C^+_{2-k-\sigma} | C^+_{1k\sigma} \rangle\rangle_\omega + (\Delta^+_{k11} + \Delta^+_{H22})\langle\langle C_{1k\sigma} | C^+_{1k\sigma} \rangle\rangle_\omega, \qquad (8)$$

$$\{\omega - (\varepsilon_{2k} - \mu + U''\langle n_1 \rangle + U'\langle n_2 \rangle)\}\langle\langle C_{2k\sigma} | C^+_{1k\sigma} \rangle\rangle_\omega = \varepsilon_{12k} \langle\langle C_{1k\sigma} | C^+_{1k\sigma} \rangle\rangle_\omega + (\Delta_{k22} + \Delta_{H11})\langle\langle C^+_{2-k-\sigma} | C^+_{1k\sigma} \rangle\rangle_\omega, \qquad (9)$$

$$\{\omega + (\varepsilon_{2k} - \mu + U''\langle n_1 \rangle + U'\langle n_2 \rangle)\}\langle\langle C^+_{2-k-\sigma} | C^+_{1k\sigma} \rangle\rangle_\omega = -\varepsilon_{12k} \langle\langle C^+_{1-k-\sigma} | C^+_{1k\sigma} \rangle\rangle_\omega + (\Delta^+_{k22} + \Delta^+_{H11})\langle\langle C_{2k\sigma} | C^+_{1k\sigma} \rangle\rangle_\omega. \qquad (10)$$

Where,

$$\eta_{1k} = \varepsilon_{1k} - \mu + U'\langle n_1 \rangle + U''\langle n_2 \rangle,$$

$$\eta_{2k} = \varepsilon_{2k} - \mu + U'\langle n_2 \rangle + U''\langle n_1 \rangle,$$

$$G_{11}(k,\omega) = \langle\langle C_{1k\sigma} | C^+_{1k\sigma} \rangle\rangle_\omega, \quad G^+_{11}(k,\omega) = \langle\langle C^+_{1-k-\sigma} | C^+_{1k\sigma} \rangle\rangle_\omega,$$

$$G_{21}(k,\omega) = \langle\langle C_{2k\sigma} | C^+_{1k\sigma} \rangle\rangle_\omega, \quad G^+_{21}(k,\omega) = \langle\langle C^+_{2-k-\sigma} | C^+_{1k\sigma} \rangle\rangle_\omega. \qquad (11)$$

And superconducting order parameters defined as

$$\Delta_{krr} = \sum_{k'} U_{kk'} \langle C_{rk'\sigma}, C_{r-k'-\sigma} \rangle, \text{ and } \Delta_{Hrr} = J_H \sum_{k'} \langle C_{rk'\sigma}, C_{r-k'-\sigma} \rangle.$$

$\langle n_1 \rangle = \sum_{k,\sigma} \langle C^+_{1k\sigma}, C_{1k\sigma} \rangle$, and $\langle n_2 \rangle = \sum_{k,\sigma} \langle C^+_{2k\sigma}, C_{2k\sigma} \rangle$ represent carrier density in two bands. In order to make the equations (7-10) tractable, we have assumed that each band have a uniform superconducting order parameter ($\Delta_{11} = \Delta_{22}$) and carrier density ($\langle n_1 \rangle = \langle n_2 \rangle = \langle n \rangle$) to avoid complexity. So simplifying the coupled equations (7-10), one can obtained the following desirable Green's functions

$$G_{11}^+(k,\omega) = \langle\langle C^+_{1-k-\sigma}, C^+_{1k\sigma} \rangle\rangle_\omega :$$

$$G_{11}^+(k,\omega) = \frac{1}{2\pi} \frac{(\Delta_k + \Delta_H)\{\omega^2 - \eta_{2k}^2 - (\Delta_k + \Delta_H)^2 - \varepsilon_{12}^2\}}{\left[\{\omega^2 - \eta_{2k}^2 - (\Delta_k + \Delta_H)^2 - \varepsilon_{12}^2\}\{\omega^2 - \eta_{1k}^2 - (\Delta_k + \Delta_H)^2 - \varepsilon_{12}^2\} - \varepsilon_{12}^2\{\eta_{1k} + \eta_{2k}\}^2\right]}. \quad (12)$$

To study the superconducting state, we have manipulated the above equation (12) in the following tractable form:

$$\langle\langle C^+_{1-k-\sigma}, C^+_{1k\sigma} \rangle\rangle_\omega = \frac{1}{2\pi}\left[\frac{B_{1k}}{(\omega - E_{1k})} + \frac{B_{2k}}{(\omega + E_{1k})} + \frac{B_{3k}}{(\omega - E_{2k})} + \frac{B_{4k}}{(\omega + E_{2k})}\right]. \quad (13)$$

Where $E_{1k}$ and $E_{2k}$ are the quasi-particle energies and ($B_{1k}, B_{2k}, B_{3k},$ and $B_{4k}$) are corresponding spectral weights in above equation (13) given as follows:

$$E_{1k} = \pm\sqrt{\frac{\{\eta_{1k}^2 + \eta_{2k}^2 + 2(\Delta_k + \Delta_H)^2 + 2\varepsilon_{12}^2\} + \{(\eta_{1k} + \eta_{2k})\sqrt{\{\eta_{2k} - \eta_{1k}\}^2 + 4\varepsilon_{12}^2}\}}{2}}, \quad (14a)$$

$$E_{2k} = \pm\sqrt{\frac{\{\eta_{1k}^2 + \eta_{2k}^2 + 2(\Delta_k + \Delta_H)^2 + 2\varepsilon_{12}^2\} - \{(\eta_{1k} + \eta_{2k})\sqrt{\{\eta_{2k} - \eta_{2k}\}^2 + 4\varepsilon_{12}^2}\}}{2}}. \quad (14b)$$

and

$$B_{1k} = \frac{\{E_{1k}^2 - \eta_{2k}^2 - (\Delta_k + \Delta_H)^2 - \varepsilon_{12}^2\}}{2E_{1k}(E_{1k}^2 - E_{2k}^2)}, \tag{15a}$$

$$B_{2k} = -B_{1k}, \tag{15b}$$

$$B_{3k} = \frac{\{E_{2k}^2 - \eta_{2k}^2 - (\Delta_k + \Delta_H)^2 - \varepsilon_{12}^2\}}{2E_{2k}(E_{2k}^2 - E_{1k}^2)}, \tag{15c}$$

$$B_{4k} = -B_{3k}. \tag{15d}$$

The superconducting state correlation $\langle C_{1-k-\sigma}^+ \cdot C_{1k\sigma}^+ \rangle$ can be calculated from the above equation (13) by using the following standard relationship:

$$\langle C_{1-k-\sigma}^+ \cdot C_{1k\sigma}^+ \rangle = i \lim_{\varepsilon \to 0} \int_{-\infty}^{\infty} \frac{\left[ \langle\langle C_{1-k-\sigma}^+ \cdot C_{1k\sigma}^+ \rangle\rangle_{\omega+i\varepsilon} - \langle\langle C_{1-k-\sigma}^+ \cdot C_{1k\sigma}^+ \rangle\rangle_{\omega-i\varepsilon} \right] \times e^{-i\omega(t-t')}}{(e^{\beta\omega}+1)} d\omega. \tag{16}$$

Where, $\beta = \frac{1}{k_B T}$, and $k_B$ is the Boltzmann constant. On simplifications and using (13) and (16), the correlation function $\langle C_{1-k-\sigma}^+ \cdot C_{1k\sigma}^+ \rangle$ representing superconducting order parameter can be obtained in desirable form as:

$$\langle C_{1-k-\sigma}^+, C_{1k\sigma}^+ \rangle = -(\Delta_k + \Delta_H)\left[ B_{1k} \tanh\left(\frac{\beta E_{1k}}{2}\right) + B_{3k} \tanh\left(\frac{\beta E_{2k}}{2}\right) \right]. \tag{17}$$

The superconducting energy gap parameter $(\Delta_k^+)$ within two band system can be written as:

$$\Delta_k^+ = \sum_k V_{kk'} \langle C_{1-k-\sigma}^+, C_{1k\sigma}^+ \rangle, \tag{18}$$

And finally, using (16) and (17) one can have:

$$\Delta_k^+ = -\sum_k V_{kk'}(\Delta_k + \Delta_H)\left[ B_{1k} \tanh\left(\frac{\beta E_{1k}}{2}\right) + B_{3k} \tanh\left(\frac{\beta E_{2k}}{2}\right) \right]. \tag{19}$$

To solve the above equation, we consider the following conditions in term of s-wave pairing symmetry ($\Delta_k = U\Delta$, $\Delta_H = J\Delta$, and $\Delta_k^+ = U\Delta$), and assume ($\Delta \to 0$ at $T \to T_c$). We also assumed the attractive interaction $V_{kk'}$ as follows:

$$V_{kk'} = U \chi(k) \chi(k') \quad if \quad |\varepsilon_k - \mu|, |\varepsilon_{k'} - \mu| \leq \omega_p$$
$$= 0, \quad otherwise,\ for\ a\ cutoff\ phonon\ energy. \tag{20}$$

Here, $\omega_p$ is the characteristic frequency of the phonon in the iron pnictide superconducting system. $\chi_{(k)}$ is the basic function for the pairing symmetry. We have avoided the electron-electron interaction in our calculations. Therefore, one can obtained an equation for $T_c$ as:

$$1 = -\frac{(U+J)}{N}\sum_k \chi^2(k)\left[B_{1k}\tanh\left(\frac{E_{1k}}{2k_B T_c}\right) + B_{3k}\tanh\left(\frac{E_{2k}}{2k_B T_c}\right)\right]. \tag{21}$$

Where $E_{1k}$ and $E_{2k}$ are the quasi-particle energies and $B_{1k}$ and $B_{3k}$ are the weights corresponding to the poles of Green's function $G_{11}^+(k,\omega) = \langle\langle C_{1-k-\sigma}^+ | C_{1k\sigma}^+\rangle\rangle_\omega$. With the help of above equation (14a & 14b) and (15a & 15c), the isotope effect coefficient ($\alpha$) can be analyzed as a function of transition temperature $T_c$. We know that phonon frequency ($\omega_p$) is inversely proportion to the isotope mass (M). Therefore, isotope coefficient can be described in the standard form as:

$$\alpha = -\frac{\partial \ln T_c}{\partial \ln M} = \frac{1}{2}\frac{\partial \ln T_c}{\partial \ln \omega_p}, \tag{22}$$

After solving equation (22), we get:

$$\alpha = \frac{1}{2}\frac{\omega_p}{T_c}\frac{\partial T_c}{\partial \omega_p}. \tag{23}$$

Further, for simplicity, we have chosen uniform orbital energy dispersions for equations (14a) and (14b). Also assumed the cylindrical Fermi surface as suggested [21,39-43] and defined as $\chi^2(k) = Cos^2(2\theta)$. Representing equation (21) as:

$$L(\omega_p) = -\frac{(U+J)}{N}Cos^2(2\theta)\sum_{rk}\left[B_{1k}\tanh\left(\frac{E_{1k}}{2k_B T_c}\right) + B_{3k}\tanh\left(\frac{E_{2k}}{2k_B T_c}\right)\right]. \tag{24}$$

We have equation (21) in terms of above function as:

$$L(\omega_p) = 1. \tag{25}$$

On differentiating above equation (25) with respect to $T_c$, we have

$$\frac{\partial L(\omega_p)}{\partial T_c} + \frac{\partial L(\omega_p)}{\partial \omega_p}\frac{\partial \omega_p}{\partial T_c} = 0. \tag{26}$$

Solving equation (23) and (26), we get a standard relation of isotope effect coefficient:

$$\alpha = -\frac{\left(\dfrac{\omega_p}{2T_c}\right)\dfrac{\partial L(\omega_p)}{\partial \omega_p}}{\dfrac{\partial L(\omega_p)}{\partial T_c}}. \tag{27}$$

Where, $L(\omega_p)$ can be written in the integral form by extending summation into integration for constant density of states at the Fermi level as follows:

$$L(\omega_p) = -\frac{1}{4\pi}\int_0^{\omega_p} d\varepsilon \int_0^{2\pi} Cos^2(2\theta)\, d\theta \left[ (B_{1k})_{\omega_p} \tanh\left(\frac{(E_{1k})_{\omega_p}}{2k_B T_c}\right) + (B_{3k})_{\omega_p} \tanh\left(\frac{(E_{2k})_{\omega_p}}{2k_B T_c}\right) \right]. \tag{28}$$

On differentiating above equation (28) with respect to $\omega_p$ and $T_c$, we have

$$\frac{\partial L(\omega_p)}{\partial \omega_p} = -\frac{1}{4\pi}\int_0^{2\pi} Cos^2(2\theta)\, d\theta \left[ (B_{1k})_{\omega_p} \tanh\left(\frac{(E_{1k})_{\omega_p}}{2k_B T_c}\right) + (B_{3k})_{\omega_p} \tanh\left(\frac{(E_{2k})_{\omega_p}}{2k_B T_c}\right) \right]. \tag{28a}$$

$$\frac{\partial L(\omega_p)}{\partial T_c} = \frac{1}{4\pi}\left(\frac{1}{2k_B T_c}\right)\int_0^{\omega_p} d\varepsilon \int_0^{2\pi} Cos^2(2\theta)\, d\theta \left[ B_{1k}\, Sech^2\left(\frac{E_{1k}}{2k_B T_c}\right) + B_{3k}\, Sech^2\left(\frac{E_{2k}}{2k_B T_c}\right) \right]. \tag{28b}$$

From equation (27), one can analyze the isotope effect coefficient as function of transition temperature $T_c$. In the next section, we have presented our numerical computation of isotope effect coefficient for two orbital per site in iron pnictide superconductors.

**Results and Discussion**

In order to analyze the influence of on-site Coulomb interactions, hybridization energy and carrier density on isotope effect coefficient, we have numerically computed the equation (27) self-consistently. We have assumed the constant density of state throughout the computation and also taken ($<n_1>$ = $<n_2>$ = $<n>$) uniform carrier density in both the band to avoid the complexity. During numerical computation, the ranges of various model parameters have been adopted from recent ARPES [41,44] and theoretical band structure calculations [19,20,22,35-37,45] relevant for iron based superconducting system.

The variation of the isotope effect coefficient as a function of $T_c$ for different values of hybridization energy ($\varepsilon_{12}$) in two orbital per site model in the iron pnictide is presented in figure 2. It is observed from figure 2 that on increasing the hybridization energy, the isotope effect coefficient increases slightly in the low temperature range. Also one can notice that the behavior of the isotope effect coefficient follows within the predicted BCS value ($\alpha$= 0.5) [4].

The variation of isotope effect coefficient ($\alpha$) as a function of $T_c$ for different value of onsite Coulomb interaction energy have been shown in figure 3. It is clear from this figure that the intraorbital Coulomb interaction suppresses the isotope effect coefficient. This decrement of isotope exponent may be explained as in terms of increase in the Coulomb interaction within two orbital per site model of the iron pnictide system. The suppression of isotope effect with U' and U" can be connected with the suppression of the spectral function, (and hence the density of electronic states) with onsite Coulomb interaction [37, 38]. Therefore, it is pointed out that the intraorbital and interorbital Coulomb interactions suppress the isotope effect coefficient in low transition range in iron based superconductors. These results are in qualitative agreement with the theoretical predictions [32, 46], where an increase in Coulomb interaction leads to decrease in the transition temperature and isotope effect exponent these systems.

In figure 4, we depict the variation of isotope effect coefficient ($\alpha$) as a function of $T_c$ for different values of carrier density (<n> = 0.05, and 0.09) in s- wave pairing symmetry (keeping parameters $\varepsilon_{12}$ = 0.0465 eV, U'= 0.07eV, U"= 0.02eV, $\omega_p$= 0.03eV fixed). Figure 4 shows that on increasing the carrier density per site in two orbital per site iron pnictide system, isotope effect coefficient ($\alpha$) exhibit large values (much higher than BCS limit [4]) at the lower temperature. While in the underdoped case, isotope coefficient shows minimum value in superconducting states in iron based systems. Further, it is important to notice that the influence of Coulomb interaction on the variation of isotope effect coefficient with transition temperature is just opposite to that of the influence of carrier density with two orbital per site model in iron based superconductors.

In summary, we have presented a BCS mean field Green's function theoretical analysis of isotope effect coefficient as a function of transition temperature by taking into account two orbitals ($d_{xz}$ & $d_{yz}$) per site model Hamiltonian applicable for iron pnictide. On the basis of numerical analysis, the influence of hybridization energy between bands, intraorbital Coulomb correlation, and carrier density on the isotope effect coefficient ($\alpha$) have been studied. It is observed that the isotope effect coefficient ($\alpha$) increases with increasing the hybridization energy ($\varepsilon_{12}$) while on increasing intraorbital Coulomb interactions, the isotope effect coefficient ($\alpha$) starts decreasing. Further, it is pointed out that the large value of the isotope effect coefficient ($\alpha$) indicates that the contribution of phonon alone is inadequate as the origin of superconductivity in doped high $T_c$ iron pnictide superconductors and need to review carefully. Therefore, it will be interesting to extend these calculations in multiorbital superconductivity for iron based superconductors and also introduced AFM ordering in the presence of Hund's energy in Fe-3$d$ orbitals. It will also be interesting that this mean field calculation of isotope effect within two orbital per site model may be extended beyond mean field (*i.e.* Dynamical mean field) approximation for better treatment of electronic correlations and spin fluctuations that exist in these iron based superconductors.

**Acknowledgement** One of the authors (Luxmi Rani) is highly thankful to Ministry of Human Resource Development (MHRD), India for financial support for this work.

**Figure Caption**

**Figure 1(a)** The Fe ions form a square lattice and unit cell contains two Fe and two As ions are shown. The As ions displaced above and below the planes of Fe is shown by doted (solid) circles respectively. **(b)** A schematic diagram of two orbital ($d_{xz}$ and $d_{yz}$) model on Fe square lattice with hopping amplitude.

**Figure 2** Isotope effect coefficients ($\alpha$) versus $T_c$ for different values of hybridization energy ($\varepsilon_{12}$) and keeping other parameters $<n>$= 0.05, U'= 0.07eV, U"= 0.02eV, $\omega_p$= 0.03eV fixed.

**Figure 3** Isotope effect coefficients ($\alpha$) versus $T_c$ for different values of onsite Coulomb interaction (U' = 0.06eV, U" = 0.02eV and U' = 0.07eV, U" = 0.02eV) and keeping other parameters $<n>$ = 0.05, $\varepsilon_{12}$ = 0.0465eV, $\omega_p$= 0.03eV fixed.

**Figure 4** Isotope effect coefficients ($\alpha$) versus $T_c$ for different values of the carrier density ($<n>$ = 0.05, and 0.09) and keeping other parameters $\varepsilon_{12}$ =0.0465 eV, U'=0.07eV, U"=0.02eV, $\omega_p$=0.03eV fixed.

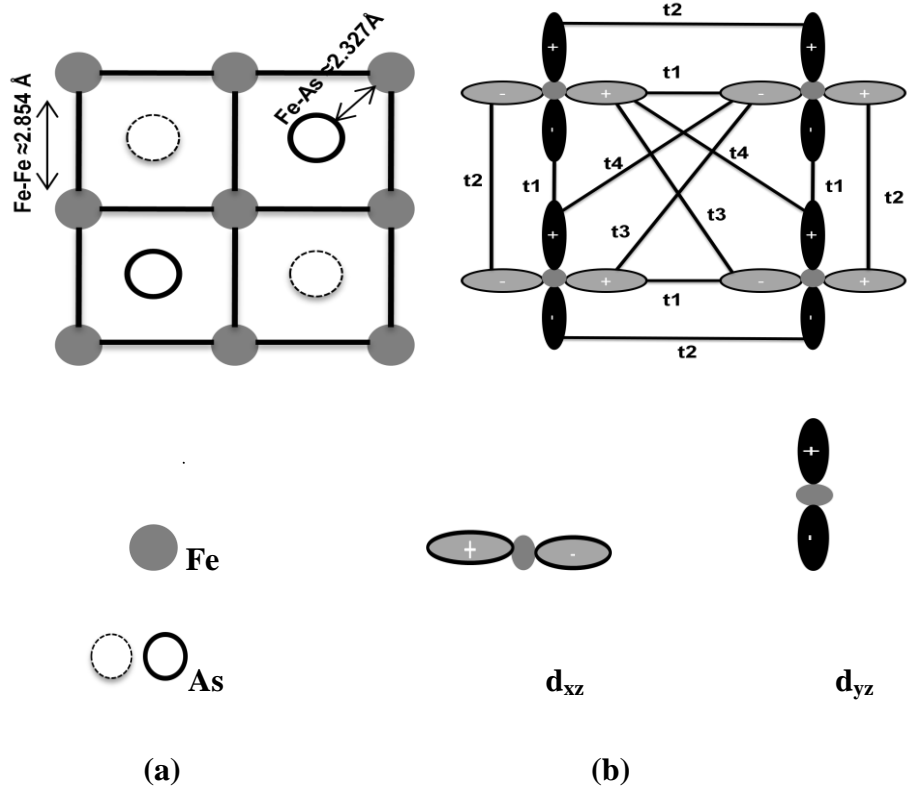

Figure 1

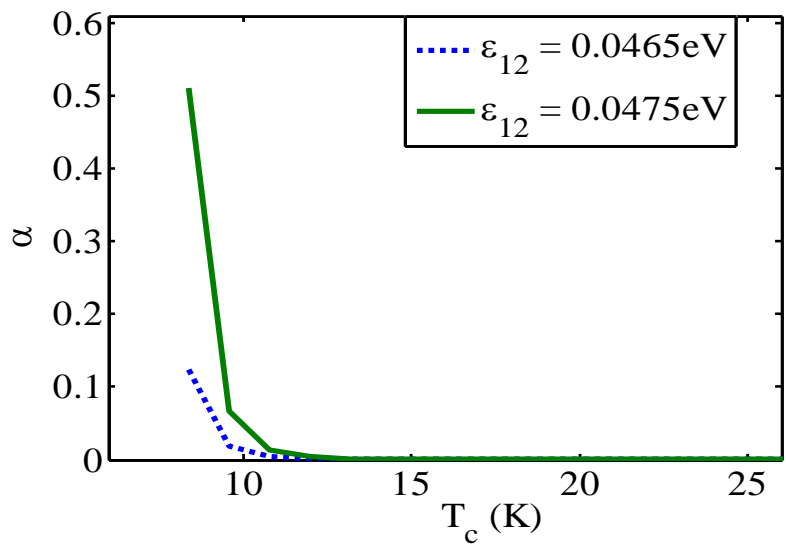

Figure 2

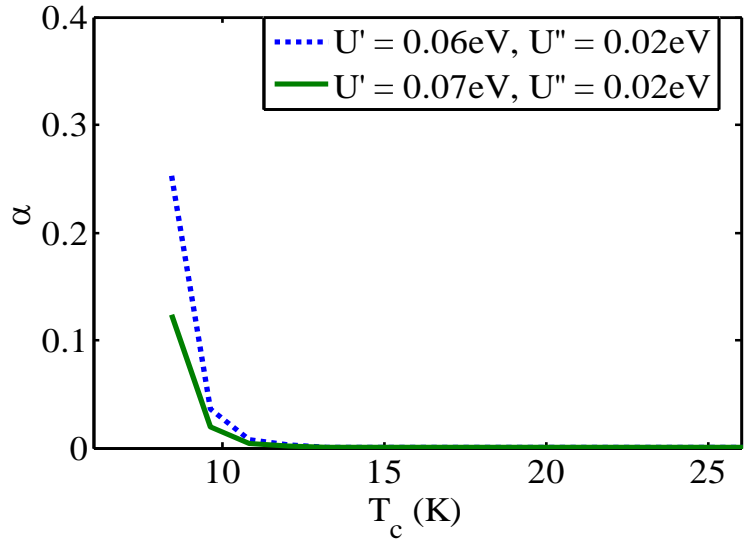

Figure 3

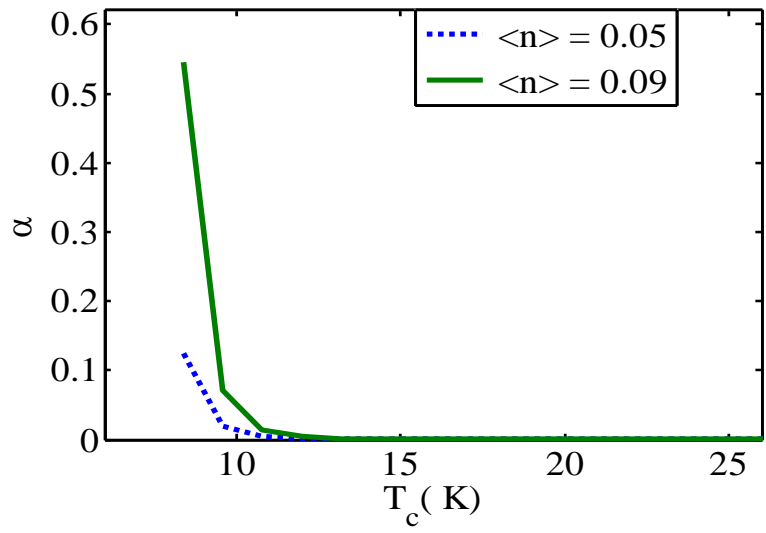

Figure 4